\documentclass[11pt]{article}
\usepackage{fullpage}
\usepackage{amsmath}
\usepackage{amssymb}
\usepackage[dvips]{epsfig}

\def\##1{{\bf{#1}}}
\def\=#1{\underline{\underline{#1}}}

\def\+#1{\underline{\bf #1}}
\def\*#1{\underline{\underline{\bf #1}}}

\def\r#1{(\ref{#1})}
\def\l#1{\label{#1}}
\def\c#1{\cite{#1}}

\def\le{\left(}
\def\ri{\right)}
\def\les{\left[}
\def\ris{\right]}
\def\lec{\left\{}
\def\ric{\right\}}

\def\.{\mbox{ \tiny{$^\bullet$} }}

\def\epso{\epsilon_{\scriptscriptstyle 0}}

\def\muo{\mu_{\scriptscriptstyle 0}}

\def\co{c_{\scriptscriptstyle 0}}

\def\eps{\epsilon}

\begin{document}

\begin{center}

{\bf {\Large Lorentz covariance of the canonical perfect lens }}

 \vspace{10mm} \large

Tom G. Mackay\footnote{Corresponding author.
E--mail: T.Mackay@ed.ac.uk}, \\
{\em School of Mathematics, University of Edinburgh, Edinburgh EH9
3JZ, UK}\\ \bigskip
 Akhlesh  Lakhtakia\footnote{ E--mail: akhlesh@psu.edu}\\{\em CATMAS~---~Computational \& Theoretical Materials Sciences Group }\\
 {\em Department of Engineering Science and
Mechanics\\ Pennsylvania State University, University Park, PA
16802--6812, USA}\\

\end{center}

\vspace{4mm}

\normalsize

\begin{abstract}
\noindent The canonical   perfect
lens~---~comprising  three slabs, each  made of  a linear,
homogeneous,  bianisotropic material with orthorhombic symmetry~---~is
Lorentz covariant.
\end{abstract}


\vspace{10mm}

\section{Introduction}
The electromagnetic properties of classical vacuum (i.e., empty
space) are characterized by its permittivity $\epso = 8.854 \times
10^{-12} \,\mbox{F m}^{-1}$ and permeability $\muo = 4 \pi \times
10^{-7}\, \mbox{H m}^{-1}$. In contrast,  \emph{anti--vacuum} has
permittivity $-\epso$ and permeability $-\muo$ \c{L02a}. The perfect lens,
 as conceptualized
by Pendry \c{Pendry01}, consists of a slab of anti--vacuum
sandwiched by two slabs of vacuum. While this perfect lens is an idealization
which can never be fully realized in practice \c{L02b}, the concept of the
perfect lens has spawned much theoretical and experimental work
within the past few years on negative refraction and metamaterials.
Indeed, interest in this area continues to escalate, with
negatively refracting metamaterials having now  entered the visible
frequency regime  \c{Dolling}. Aside from metamaterials,  negative
refraction occurs in biological scenarios \c{Stavenga}, and there is
the possibility of negative refraction arising in special \c{ML04}
and general \c{LMS05} relativistic scenarios.

 A
fundamental characteristic of  vacuum  is that its constitutive
parameters are invariant under a Lorentz transformation. A
straightforward derivation reveals that the constitutive parameters
of anti--vacuum are also invariant under a Lorentz transformation.
Therefore, the vacuum/anti--vacuum/vacuum perfect lens is Lorentz
covariant. A canonical formulation for the perfect lens has also
been developed, wherein
the two constituent materials are linear, homogeneous,
orthorhombic materials \c{L02b,LS03}. In this Letter, we
address the question: is this canonical perfect lens invariant under
a Lorentz transformation?

\section{Canonical perfect lens}

The canonical perfect  lens comprises a slab of material
labeled $b$, sandwiched between two slabs of a material labeled $a$, as
schematically illustrated in Figure~\ref{fig1}. With respect to an inertial frame of reference $\Sigma$, material
 $a$
occupies the regions $0 \leq z < d_1$ and $d_1 + d_2  \leq z < d_1 +
d_2 +  d_3$, while material $b$ occupies the region $d_1   \leq z <
d_1 + d_2 $. Both materials move with a common and uniform velocity
$\#v=v\hat{\#x}$, where $v \in (-\co, \co)$ and $\co$ is the speed of
light in vacuum; thus, the
direction of relative motion is parallel to the interfaces between
material $a$ and material $b$.

The materials $a$ and $b$ are linear and homogeneous. With respect
to an inertial reference frame $\tilde{\Sigma}$ that also moves
at velocity $\#v$ relative to $\Sigma$, their
frequency--domain constitutive relations are
\begin{equation}
\left. \begin{array}{l} \tilde{ \#D} ( \tilde{x}, \tilde{y},
\tilde{z}, \tilde{\omega}) = \epso \les \tilde{\=\eps}^{a,b} \.
\tilde{\#E} ( \tilde{x}, \tilde{y}, \tilde{z}, \tilde{\omega}) +
\tilde{ \=\alpha}^{a,b} \. \tilde{\#H} ( \tilde{x}, \tilde{y},
\tilde{z}, \tilde{\omega}) \ris \vspace{6pt} \\
\tilde{ \#B} ( \tilde{x}, \tilde{y}, \tilde{z}, \tilde{\omega}) =
\muo \les \tilde{\=\beta}^{a,b} \.\tilde{ \#E }( \tilde{x},
\tilde{y}, \tilde{z}, \tilde{\omega}) +
 \tilde{\=\mu}^{a,b} \. \tilde{\#H }( \tilde{x}, \tilde{y},
\tilde{z}, \tilde{\omega}) \ris \l{CR_1}
\end{array}
\right\},
\end{equation}
wherein the 3$\times$3 constitutive dyadics have the orthorhombic
form
\begin{equation}
\tilde{\=\xi}^{a,b} = \les \begin{array}{ccc} \tilde{\xi}^{a,b}_{11}
& 0 & 0 \\ 0&  \tilde{\xi}^{a,b}_{22} & 0 \\ 0 & 0 &
\tilde{\xi}^{a,b}_{33} \end{array} \ris, \qquad \qquad \le \xi =
\eps, \alpha, \beta, \mu \ri.
\label{aaa}
\end{equation}
 With respect to $\Sigma$, their
frequency--domain constitutive relations
are given as \c{Chen}
\begin{equation}
\left. \begin{array}{l}  \#D ( x, y, z, \omega) = \epso \les
\=\eps^{a,b} \. \#E ( x, y, z, \omega) + \=\alpha^{a,b} \. \#H ( x, y, z, \omega)
 \ris \vspace{6pt} \\
 \#B ( x, y, z, \omega) = \muo
\les \=\beta^{a,b} \. \#E ( x, y, z, \omega) +
 \=\mu^{a,b} \. \#H ( x, y, z, \omega) \ris \l{CR_2}
\end{array}
\right\},
\end{equation}
wherein the 3$\times$3 constitutive dyadics have the form
\begin{equation}
\=\xi^{a,b} = \les \begin{array}{ccc} \xi^{a,b}_{11} & 0 & 0
\\ 0&  \xi^{a,b}_{22} & \xi^{a,b}_{23} \\ 0 & \xi^{a,b}_{32} & \xi^{a,b}_{33}
\end{array} \ris, \qquad \qquad \le \xi = \eps, \alpha, \beta, \mu
\ri.
\end{equation}
Explicit expressions for the  components of $\=\xi^{a,b}$, in terms
of $v$ and the components of $\={\tilde \xi}^{a,b}$, are provided in
Appendix~1.

\section{Lorentz covariance}

Following the approach developed for the canonical perfect lens in the
co--moving reference frame $\tilde\Sigma$ \c{L02b,LS03}, we express the
electromagnetic phasors $\#E (x,y,z,\omega)$, etc., in $\Sigma$    in terms of their spatial Fourier transformations
with respect to $x$ and $y$; thus,
\begin{equation}
\left.
\begin{array}{l}
\#E (x,y,z,\omega) = \#e (z, \kappa, \psi, \omega) \, \exp \les i
\kappa \le x \cos \psi + y \sin \psi \ri \ris \vspace{4pt}\\
\#B (x,y,z,\omega) = \#b (z, \kappa, \psi, \omega) \, \exp \les i
\kappa \le x \cos \psi + y \sin \psi \ri \ris\vspace{4pt}\\
\#D (x,y,z,\omega) = \#d (z, \kappa, \psi, \omega) \, \exp \les i
\kappa \le x \cos \psi + y \sin \psi \ri \ris \vspace{4pt}\\
\#H (x,y,z,\omega) = \#h (z, \kappa, \psi, \omega) \, \exp \les i
\kappa \le x \cos \psi + y \sin \psi \ri \ris
\end{array}
\right\}.
\end{equation}
Thereby, wave propagation in the non--co--moving reference frame is
described by the 4$\times$4 matrix ordinary differential equations
\begin{eqnarray}
\nonumber
&&
\frac{d}{dz} \les \underline{f} (z, \kappa, \psi, \omega) \ris = i
\les \=P_{\,a} ( \kappa, \psi, \omega) \ris \. \les \underline{f}
(z, \kappa, \psi, \omega) \ris\,, \\
&&\qquad\qquad\quad z\in(0,d_1)\, {\rm or}
\,
z\in (d_1+d_2,d_1+d_2+d_3)\,,
\l{MODEa}
\end{eqnarray}
and
\begin{eqnarray}
\nonumber
&&
\frac{d}{dz} \les \underline{f} (z, \kappa, \psi, \omega) \ris = i
\les \=P_{\,b} ( \kappa, \psi, \omega) \ris \. \les \underline{f}
(z, \kappa, \psi, \omega) \ris\,, \\
&&\qquad\qquad\quad z\in (d_1,d_1+d_2)\,,
\l{MODEb}
\end{eqnarray}
with the column 4--vector
\begin{equation}
\les \,\underline{f} \, \ris = \les \#e \. \hat{\#x},\, \#e \.
\hat{\#y}, \,\#h \. \hat{\#x}, \,\#h \. \hat{\#y} \ris^T.
\end{equation}
Explicit expressions for the  components of the 4$\times$4 matrixes
$ \les \=P_{\,a,b} \ris $ are provided in Appendix~2.

By solving
\r{MODEa} and \r{MODEb}, we see that the phasors at $z=0$ and $z = d_1 + d_2 + d_3$
are related as
\begin{eqnarray}
 \les
\,\underline{f} (d_1 + d_2 + d_3, \kappa, \psi, \omega ) \, \ris &=&
\exp \lec i d_3 \les \=P_{\,a} (\kappa, \psi, \omega) \ris \ric \.
\exp \lec i d_2 \les \=P_{\,b} (\kappa, \psi, \omega) \ris \ric
\nonumber \\ && \. \exp \lec i d_1 \les \=P_{\,a} (\kappa, \psi,
\omega) \ris \ric
 \. \les \,\underline{f} (0, \kappa, \psi, \omega )
\, \ris. \l{MODE_sol}
\end{eqnarray}
As described  elsewhere \c{L02b,LS03}, the solution of the problem
of the canonical perfect lens
involves finding the thicknesses $d_1$ and   $d_3$ for
material $a$, and the thickness $d_2$ for material $b$, such that
\begin{equation}
\les \,\underline{f} (0, \kappa, \psi, \omega ) \, \ris = \les
\,\underline{f} (d_1 + d_2 + d_3, \kappa, \psi, \omega ) \, \ris
\l{f_cond}
\end{equation}
for all $\kappa$, $\psi$ and $\omega$.

The apparently simplest route
to satisfying the perfect-lens condition \r{f_cond} is to ensure that
the matrixes $\=P_{\,a} (\kappa, \psi, \omega)$ and
$\=P_{\,b} (\kappa, \psi, \omega)$ commute for all $\kappa$, $\psi$ and $\omega$.
Then,
\begin{equation}
\=P_{\,b} (\kappa, \psi, \omega)  + \gamma \=P_{\,a} (\kappa, \psi,
\omega) = \=0\,, \l{commute}
\end{equation}
and
\begin{equation}
d_1 + d_3 = \gamma d_2\,, \l{d_cond}
\end{equation}
where $\gamma>0$ is some scalar.
A straightforward calculation reveals that \r{commute} holds for the
reference frame $\Sigma$  when
\begin{equation}
\left.
\begin{array}{l}
\tilde{\xi}^b_{11} + \gamma \tilde{\xi}^a_{11} = 0 \vspace{4pt} \\
\tilde{\xi}^b_{22} + \gamma \tilde{\xi}^a_{22} = 0 \vspace{4pt} \\
\tilde{\xi}^b_{33} + \gamma^{-1} \tilde{\xi}^a_{33 } = 0
\end{array}
\right\}, \qquad \qquad (\xi = \eps, \alpha, \beta, \mu) .
\l{xi_cond}
\end{equation}
In particular,  since the conditions \r{d_cond} and \r{xi_cond} hold
for arbitrary $v \in (-\co, \co)$, the canonical perfect lens
is Lorentz covariant.

Thus, not only is a perfect lens comprising slabs of vacuum,
anti--vacuum, and vacuum   Lorentz covariant, but combinations of
linear, homogeneous, and orthorhombic mediums $a$ and $b$ can be
found such that a perfect lens made thereof is also Lorentz
covariant. The consequences of this result for space exploration and
observational astronomy are matters for future consideration.

\vspace{20mm}

\noindent{\bf Acknowledgement:} TGM is supported by a \emph{Royal
Society of Edinburgh/Scottish Executive Support Research
Fellowship}.\\

\section*{Appendix~1}

The components of the 3$\times$3 constitutive dyadics  in $\Sigma$
are provided by a straightforward, but cumbersome, application of
the Lorentz transformation to the constitutive dyadics
in $\tilde\Sigma$ \c{Chen}. Thus,
\begin{eqnarray}
\eps^{a,b}_{11} &=& \tilde{\eps}^{a,b}_{11},\\
\eps^{a,b}_{22} &=& - \frac{ \co^2 - v^2 }{\Delta}  \les \co^2
\tilde{\eps}^{a,b}_{22} + v^2 \tilde{\eps}^{a,b}_{33} \le
\tilde{\alpha}^{a,b}_{22} \tilde{\beta}^{a,b}_{22} -
\tilde{\eps}^{a,b}_{22} \tilde{\mu}^{a,b}_{22} \ri \ris
\tilde{\mu}^{a,b}_{22} \tilde{\mu}^{a,b}_{33},
\\
\eps^{a,b}_{23} &=& - \frac{  \co^2 - v^2}{\Delta} \le
\tilde{\eps}^{a,b}_{33} \tilde{\alpha}^{a,b}_{22} \epso -
\tilde{\eps}^{a,b}_{22} \tilde{\beta}^{a,b}_{33} \muo \ri v \co^2
\tilde{\mu}^{a,b}_{22} \tilde{\mu}^{a,b}_{33},
\\
\eps^{a,b}_{32} &=& \frac{ \co^2 - v^2  }{\Delta} \le
\tilde{\eps}^{a,b}_{22} \tilde{\alpha}^{a,b}_{33} \epso -
\tilde{\eps}^{a,b}_{33} \tilde{\beta}^{a,b}_{22} \muo \ri v \co^2
\tilde{\mu}^{a,b}_{22} \tilde{\mu}^{a,b}_{33} ,
\\
\eps^{a,b}_{33} &=& - \frac{ \co^2 - v^2 }{\Delta}
 \les \co^2
\tilde{\eps}^{a,b}_{33} + v^2 \tilde{\eps}^{a,b}_{22} \le
\tilde{\alpha}^{a,b}_{33} \tilde{\beta}^{a,b}_{33} -
\tilde{\eps}^{a,b}_{33} \tilde{\mu}^{a,b}_{33} \ri \ris
\tilde{\mu}^{a,b}_{22} \tilde{\mu}^{a,b}_{33},
\end{eqnarray}

\begin{eqnarray}
\alpha^{a,b}_{11} &=& \tilde{\alpha}^{a,b}_{11},\\
\alpha^{a,b}_{22} &=& - \frac{ \co^2 - v^2 }{\Delta}
 \les \co^2
\tilde{\alpha}^{a,b}_{22} \epso + v^2 \tilde{\beta}^{a,b}_{33} \muo
\le \tilde{\alpha}^{a,b}_{22} \tilde{\beta}^{a,b}_{22} -
\tilde{\eps}^{a,b}_{22} \tilde{\mu}^{a,b}_{22} \ri \ris
\tilde{\mu}^{a,b}_{22} \tilde{\mu}^{a,b}_{33} \muo \co^2,
\\
\alpha^{a,b}_{23} &=& - \frac{ v  \co^4 \tilde{\mu}^{a,b}_{22}
\tilde{\mu}^{a,b}_{33}}{\Delta} \Bigg( \muo \le 1 -
\tilde{\eps}^{a,b}_{22} \tilde{\mu}^{a,b}_{33} + \muo^2 v^2
\tilde{\beta}^{a,b}_{22} \tilde{\beta}^{a,b}_{33} \ri + \epso \Big\{
\tilde{\alpha}^{a,b}_{22} \tilde{\alpha}^{a,b}_{33}\nonumber
\\&&  + v^2 \les
\tilde{\alpha}^{a,b}_{22} \tilde{\beta}^{a,b}_{22} \le
\tilde{\alpha}^{a,b}_{33} \tilde{\beta}^{a,b}_{33} -
\tilde{\eps}^{a,b}_{33} \tilde{\mu}^{a,b}_{33} \ri-
\tilde{\mu}^{a,b}_{22} \muo^2\le \tilde{\alpha}^{a,b}_{33}
\tilde{\beta}^{a,b}_{33} \tilde{\eps}^{a,b}_{22}+
\tilde{\eps}^{a,b}_{33} - \tilde{\eps}^{a,b}_{22}
\tilde{\eps}^{a,b}_{33} \tilde{\mu}^{a,b}_{33}
 \ri
\ris \Big\} \Bigg) ,
\\
\alpha^{a,b}_{32} &=& \frac{ v  \co^4 \tilde{\mu}^{a,b}_{22}
\tilde{\mu}^{a,b}_{33}}{\Delta} \Bigg( \muo \le 1 -
\tilde{\eps}^{a,b}_{33} \tilde{\mu}^{a,b}_{22} + \muo^2 v^2
\tilde{\beta}^{a,b}_{22} \tilde{\beta}^{a,b}_{33} \ri + \epso \Big\{
\tilde{\alpha}^{a,b}_{22} \tilde{\alpha}^{a,b}_{33}\nonumber
\\&&  + v^2 \les
\tilde{\alpha}^{a,b}_{22} \tilde{\beta}^{a,b}_{22} \le
\tilde{\alpha}^{a,b}_{33} \tilde{\beta}^{a,b}_{33} -
\tilde{\eps}^{a,b}_{33} \tilde{\mu}^{a,b}_{33} \ri-
\tilde{\eps}^{a,b}_{22} \muo^2\le \tilde{\alpha}^{a,b}_{33}
\tilde{\beta}^{a,b}_{33} \tilde{\mu}^{a,b}_{22}+
\tilde{\mu}^{a,b}_{33} - \tilde{\eps}^{a,b}_{33}
\tilde{\mu}^{a,b}_{22} \tilde{\mu}^{a,b}_{33}
 \ri
\ris \Big\} \Bigg) ,
\\
\alpha^{a,b}_{33} &=& - \frac{ \co^2 - v^2 }{\Delta}
 \les \co^2
\tilde{\alpha}^{a,b}_{33} \epso + v^2 \tilde{\beta}^{a,b}_{22} \muo
\le \tilde{\alpha}^{a,b}_{33} \tilde{\beta}^{a,b}_{33} -
\tilde{\eps}^{a,b}_{33} \tilde{\mu}^{a,b}_{33} \ri \ris
\tilde{\mu}^{a,b}_{22} \tilde{\mu}^{a,b}_{33} \muo \co^2,
\end{eqnarray}

\begin{eqnarray}
\beta^{a,b}_{11} &=& \tilde{\beta}^{a,b}_{11},\\
\beta^{a,b}_{22} &=& - \frac{ \co^2 - v^2 }{\Delta}
 \les \co^2
\tilde{\beta}^{a,b}_{22} \muo + v^2 \tilde{\alpha}^{a,b}_{33} \epso
\le \tilde{\alpha}^{a,b}_{22} \tilde{\beta}^{a,b}_{22} -
\tilde{\eps}^{a,b}_{22} \tilde{\mu}^{a,b}_{22} \ri \ris
\tilde{\mu}^{a,b}_{22} \tilde{\mu}^{a,b}_{33} \epso \co^2,
\\
\beta^{a,b}_{23} &=& - \frac{ v  \co^4 \tilde{\mu}^{a,b}_{22}
\tilde{\mu}^{a,b}_{33}}{\Delta} \Bigg[ \muo \tilde{\beta}^{a,b}_{22}
\tilde{\beta}^{a,b}_{33} + \epso \Big( 1 - \tilde{\eps}^{a,b}_{33}
\tilde{\mu}^{a,b}_{22} + \epso v^2 \Big\{ \epso
\tilde{\alpha}^{a,b}_{22} \tilde{\alpha}^{a,b}_{33} \nonumber \\ &&
+ \muo \les \tilde{\alpha}^{a,b}_{22} \tilde{\beta}^{a,b}_{22}
 \le
 \tilde{\alpha}^{a,b}_{33}
\tilde{\beta}^{a,b}_{33} - \tilde{\eps}^{a,b}_{33}
\tilde{\mu}^{a,b}_{33} \ri - \tilde{\eps}^{a,b}_{22} \le
\tilde{\alpha}^{a,b}_{33} \tilde{\beta}^{a,b}_{33}
\tilde{\mu}^{a,b}_{22} + \tilde{\mu}^{a,b}_{33}-
\tilde{\eps}^{a,b}_{33}\tilde{\mu}^{a,b}_{22}\tilde{\mu}^{a,b}_{33}
\ri \ris
 \Big\} \Big) \Bigg] ,
\\
\beta^{a,b}_{32} &=&  \frac{ v  \co^4 \tilde{\mu}^{a,b}_{22}
\tilde{\mu}^{a,b}_{33}}{\Delta} \Bigg[ \muo \tilde{\beta}^{a,b}_{22}
\tilde{\beta}^{a,b}_{33} + \epso \Big( 1 - \tilde{\eps}^{a,b}_{22}
\tilde{\mu}^{a,b}_{33} + \epso v^2 \Big\{ \epso
\tilde{\alpha}^{a,b}_{22} \tilde{\alpha}^{a,b}_{33} \nonumber \\ &&
+ \muo \les \tilde{\alpha}^{a,b}_{22} \tilde{\beta}^{a,b}_{22}
 \le
 \tilde{\alpha}^{a,b}_{33}
\tilde{\beta}^{a,b}_{33} - \tilde{\eps}^{a,b}_{33}
\tilde{\mu}^{a,b}_{33} \ri - \tilde{\mu}^{a,b}_{22} \le
\tilde{\alpha}^{a,b}_{33} \tilde{\beta}^{a,b}_{33}
\tilde{\eps}^{a,b}_{22} + \tilde{\eps}^{a,b}_{33}-
\tilde{\eps}^{a,b}_{22}\tilde{\eps}^{a,b}_{33}\tilde{\mu}^{a,b}_{33}
\ri \ris
 \Big\} \Big) \Bigg] ,
\\
\beta^{a,b}_{33} &=& - \frac{ \co^2 - v^2 }{\Delta}
 \les \co^2
\tilde{\beta}^{a,b}_{33} \muo + v^2 \tilde{\alpha}^{a,b}_{22} \epso
\le \tilde{\alpha}^{a,b}_{33} \tilde{\beta}^{a,b}_{33} -
\tilde{\eps}^{a,b}_{33} \tilde{\mu}^{a,b}_{33} \ri \ris
\tilde{\mu}^{a,b}_{22} \tilde{\mu}^{a,b}_{33} \epso \co^2,
\end{eqnarray}

\begin{eqnarray}
\mu^{a,b}_{11} &=& \tilde{\mu}^{a,b}_{11},\\
\mu^{a,b}_{22} &=& - \frac{ \co^2 - v^2 }{\Delta}  \les \co^2
\tilde{\mu}^{a,b}_{22} + v^2 \tilde{\mu}^{a,b}_{33} \le
\tilde{\alpha}^{a,b}_{22} \tilde{\beta}^{a,b}_{22} -
\tilde{\eps}^{a,b}_{22} \tilde{\mu}^{a,b}_{22} \ri \ris
\tilde{\mu}^{a,b}_{22} \tilde{\mu}^{a,b}_{33},
\\
\mu^{a,b}_{23} &=& - \frac{ \co^2 - v^2 }{\Delta} \le
\tilde{\alpha}^{a,b}_{33} \tilde{\mu}^{a,b}_{22} \epso -
\tilde{\beta}^{a,b}_{22} \tilde{\mu}^{a,b}_{33} \muo \ri  v \co^2
\tilde{\mu}^{a,b}_{22} \tilde{\mu}^{a,b}_{33},
\\
\mu^{a,b}_{32} &=& \frac{ \co^2 - v^2  }{\Delta} \le
\tilde{\alpha}^{a,b}_{22} \tilde{\beta}^{a,b}_{33} \epso -
\tilde{\beta}^{a,b}_{33} \tilde{\mu}^{a,b}_{22} \muo \ri v \co^2
\tilde{\mu}^{a,b}_{22} \tilde{\mu}^{a,b}_{33},
\\
\mu^{a,b}_{33} &=& - \frac{ \co^2 - v^2 }{\Delta}
 \les \co^2
\tilde{\mu}^{a,b}_{33} + v^2 \tilde{\mu}^{a,b}_{22} \le
\tilde{\alpha}^{a,b}_{33} \tilde{\beta}^{a,b}_{33} -
\tilde{\eps}^{a,b}_{33} \tilde{\mu}^{a,b}_{33} \ri \ris
\tilde{\mu}^{a,b}_{22} \tilde{\mu}^{a,b}_{33},
\end{eqnarray}
where
\begin{eqnarray}
\Delta &=& \co^4 v^2 \le \tilde{\alpha}^{a,b}_{22}
\tilde{\mu}^{a,b}_{33} \epso - \tilde{\beta}^{a,b}_{33}
\tilde{\mu}^{a,b}_{22} \muo \ri \le \tilde{\beta}^{a,b}_{22}
\tilde{\mu}^{a,b}_{33} \muo - \tilde{\alpha}^{a,b}_{33}
\tilde{\mu}^{a,b}_{22} \epso \ri \nonumber \\ && - \les \ \co^2
\tilde{\mu}^{a,b}_{22} + v^2 \tilde{\mu}^{a,b}_{33} \le
\tilde{\alpha}^{a,b}_{22} \tilde{\beta}^{a,b}_{22}-
\tilde{\eps}^{a,b}_{22} \tilde{\mu}^{a,b}_{22} \ri \ris
\nonumber
\\
&&\quad\times\les \ \co^2
\tilde{\mu}^{a,b}_{33} + v^2 \tilde{\mu}^{a,b}_{22} \le
\tilde{\alpha}^{a,b}_{33} \tilde{\beta}^{a,b}_{33}-
\tilde{\eps}^{a,b}_{33} \tilde{\mu}^{a,b}_{33} \ri \ris.
\end{eqnarray}

\section*{Appendix~2}

The matrix ordinary differential equation approach to solving two--point
boundary--value
problems, as implemented for this Letter,  is described at length elsewhere
\c{LWmap}. The components of the 4$\times$4 matrixes $ \les
\=P_{\,a,b} \ris $ are delivered by a straightforward manipulation
of the frequency--domain
Maxwell postulates, together with the constitutive relations
\r{CR_2}. Thus,
\begin{eqnarray}
 \les \=P_{\,a,b} \ris_{11} &=& \frac{\kappa_y}{\rho^{a,b}} \le
 \frac{\kappa_x \alpha^{a,b}_{33}}{ \omega \muo} + \alpha^{a,b}_{33} \beta^{a,b}_{23}
 - \eps^{a,b}_{33} \mu^{a,b}_{23} \ri,
\\ \les \=P_{\,a,b} \ris_{12} &=& \omega \muo \beta^{a,b}_{22} -
\frac{1}{\omega \rho^{a,b}} \Bigg( \alpha^{a,b}_{33} \les
\frac{1}{\muo} \kappa^2_x + \omega \kappa_x  \le \beta^{a,b}_{23} -
\beta^{a,b}_{32} \ri - \omega^2 \muo \beta^{a,b}_{23}
\beta^{a,b}_{33} \ris \nonumber \\ &&
  - \omega \lec \mu^{a,b}_{23} \les \omega \muo \eps^{a,b}_{32}
\beta^{a,b}_{33} + \eps^{a,b}_{33} \le \kappa_x - \omega  \muo
\beta^{a,b}_{32} \ri \ris - \eps^{a,b}_{32} \mu^{a,b}_{33} \le
\kappa_x + \omega \muo \beta^{a,b}_{23} \ri
\ric \Bigg),\\
 \les \=P_{\,a,b} \ris_{13} &=&
\frac{\kappa_y}{\rho^{a,b}} \les
 \frac{\kappa_x \mu^{a,b}_{33}}{ \omega \epso} + \frac{ \muo \le \mu^{a,b}_{33} \beta^{a,b}_{23}
 - \beta^{a,b}_{33} \mu^{a,b}_{23} \ri}{\epso} \ris,\\
 \les \=P_{\,a,b} \ris_{14} &=& \omega \muo \mu^{a,b}_{22} -
\frac{1}{\omega \rho^{a,b}} \Bigg( \mu^{a,b}_{33} \les
\frac{1}{\epso} \kappa^2_x + \omega \kappa_x  \le \alpha^{a,b}_{32}
+ \frac{\muo}{\epso} \beta^{a,b}_{23} \ri + \omega^2 \muo
\alpha^{a,b}_{32} \beta^{a,b}_{23} \ris \nonumber
\\ &&
  + \omega \lec \mu^{a,b}_{23} \les \omega \muo \eps^{a,b}_{33}
\beta^{a,b}_{32} - \beta^{a,b}_{33} \le \frac{\muo}{\epso} \kappa_x
+ \omega \muo \alpha^{a,b}_{32} \ri \ris - \eps^{a,b}_{33}
\mu^{a,b}_{32} \le \kappa_x + \omega \muo \beta^{a,b}_{23} \ri \ric
\Bigg), \nonumber \\ &&
\end{eqnarray}
\begin{eqnarray}
 \les \=P_{\,a,b} \ris_{21} &=&
-\omega \muo \beta^{a,b}_{11} +
\frac{\kappa^2_y \alpha^{a,b}_{33}}{\omega \muo \rho^{a,b}},\\
 \les \=P_{\,a,b} \ris_{22} &=& - \frac{\kappa_y}{\rho^{a,b}} \le
 \frac{\kappa_x \alpha^{a,b}_{33}}{ \omega \muo} - \alpha^{a,b}_{33} \beta^{a,b}_{32}
 + \eps^{a,b}_{32} \mu^{a,b}_{33} \ri,\\
 \les \=P_{\,a,b} \ris_{23} &=& -\omega \muo \mu^{a,b}_{11} +
\frac{\kappa^2_y \mu^{a,b}_{33}}{\omega \epso \rho^{a,b}},\\\les
\=P_{\,a,b} \ris_{24} &=& - \frac{\kappa_y}{\rho^{a,b}} \le
 \frac{\kappa_x \mu^{a,b}_{33}}{ \omega \epso} - \alpha^{a,b}_{33} \mu^{a,b}_{32}
 + \alpha^{a,b}_{32} \mu^{a,b}_{33} \ri,
 \end{eqnarray}
\begin{eqnarray}
  \les \=P_{\,a,b} \ris_{31} &=& - \frac{\kappa_y}{\rho^{a,b}} \les
 \frac{\kappa_x \eps^{a,b}_{33}}{ \omega \muo} + \frac{\epso \le \alpha^{a,b}_{33} \eps^{a,b}_{23}
 - \eps^{a,b}_{33} \alpha^{a,b}_{23} \ri}{ \muo} \ris,\\
 \les \=P_{\,a,b} \ris_{32} &=& - \omega \epso \eps^{a,b}_{22} +
\frac{1}{\omega \rho^{a,b}} \Bigg( \eps^{a,b}_{33} \les
\frac{1}{\muo} \kappa^2_x - \omega \kappa_x  \le \frac{\epso}{\muo}
\alpha^{a,b}_{23} + \beta^{a,b}_{32} \ri + \omega^2 \epso
\alpha^{a,b}_{23} \beta^{a,b}_{32} \ris \nonumber
\\ &&
  + \omega \lec \eps^{a,b}_{23} \les \omega \epso \eps^{a,b}_{32}
\mu^{a,b}_{33} + \alpha^{a,b}_{33} \le \frac{\epso}{\muo} \kappa_x -
\omega \epso \beta^{a,b}_{32} \ri \ris + \eps^{a,b}_{32}
\beta^{a,b}_{33} \le \kappa_x
-\omega \epso \alpha^{a,b}_{23} \ri \ric \Bigg),\\
 \les \=P_{\,a,b} \ris_{33} &=& - \frac{\kappa_y}{\rho^{a,b}} \le
 \frac{\kappa_x \beta^{a,b}_{33}}{ \omega \epso} - \alpha^{a,b}_{23} \beta^{a,b}_{33}
 + \eps^{a,b}_{23} \mu^{a,b}_{33} \ri,\\
  \les \=P_{\,a,b} \ris_{34} &=& -\omega \epso \alpha^{a,b}_{22} +
\frac{1}{\omega \rho^{a,b}} \Bigg( \beta^{a,b}_{33} \les
\frac{1}{\epso} \kappa^2_x + \omega \kappa_x  \le \alpha^{a,b}_{32}
- \alpha^{a,b}_{23} \ri - \omega^2 \epso \alpha^{a,b}_{23}
\alpha^{a,b}_{32} \ris \nonumber \\ &&
  - \omega \lec \mu^{a,b}_{32} \les \omega \epso
\eps^{a,b}_{23} \alpha^{a,b}_{33} + \eps^{a,b}_{33} \le \kappa_x -
\omega \epso \alpha^{a,b}_{23} \ri \ris + \eps^{a,b}_{23}
\mu^{a,b}_{33} \le \kappa_x + \omega \epso \alpha^{a,b}_{32} \ri
\ric \Bigg),
\end{eqnarray}
\begin{eqnarray}
 \les \=P_{\,a,b} \ris_{41} &=&
\omega \epso \eps^{a,b}_{11} -
\frac{\kappa^2_y \eps^{a,b}_{33}}{\omega \muo \rho^{a,b}},\\
 \les \=P_{\,a,b} \ris_{42} &=& \frac{\kappa_y}{\rho^{a,b}} \le
 \frac{\kappa_x \eps^{a,b}_{33}}{ \omega \muo} - \eps^{a,b}_{33} \beta^{a,b}_{32}
 + \eps^{a,b}_{32} \beta^{a,b}_{33} \ri,\\
 \les \=P_{\,a,b} \ris_{43} &=& \omega \epso \alpha^{a,b}_{11} -
\frac{\kappa^2_y \beta^{a,b}_{33}}{\omega \epso \rho^{a,b}},
\\\les \=P_{\,a,b}
\ris_{44} &=& \frac{\kappa_y}{\rho^{a,b}} \le
 \frac{\kappa_x \beta^{a,b}_{33}}{ \omega \epso} + \alpha^{a,b}_{32} \beta^{a,b}_{33}
 - \eps^{a,b}_{33} \mu^{a,b}_{32} \ri,
\end{eqnarray}
with
\begin{equation}
\left.
\begin{array}{c}
\rho^{a,b} = \eps^{a,b}_{33} \mu^{a,b}_{33} - \alpha^{a,b}_{33}
\beta^{a,b}_{33}\\[4pt]
\kappa_x = \kappa \cos \psi\\[4pt]
\kappa_y = \kappa \sin \psi
\end{array}\right\}\,.
\end{equation}

\vspace{25mm}

 \setcounter{figure}{0}
\begin{figure}[!ht]
\centering \psfull \epsfig{file=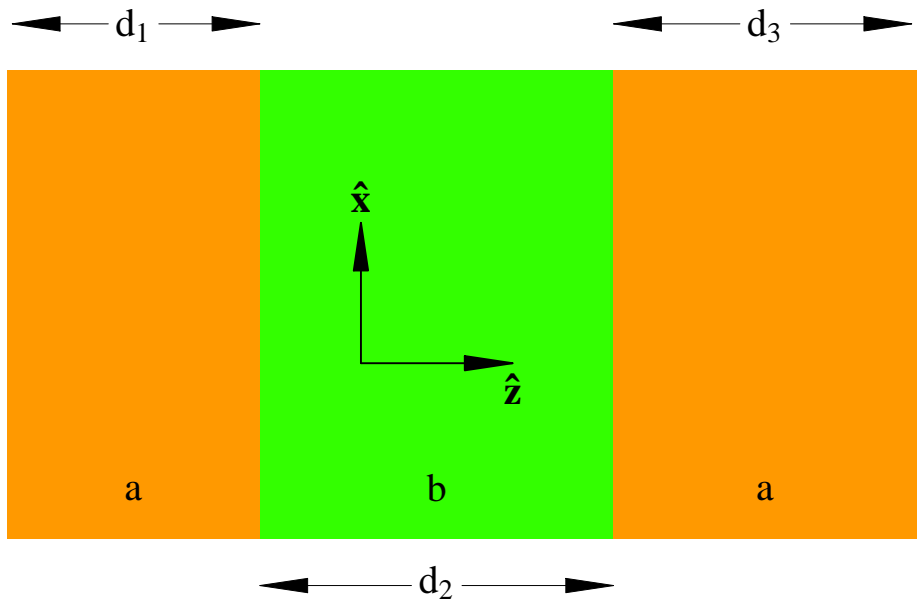,width=5.0in}
 \caption{\label{fig1}
Schematic of the canonical perfect lens.
 }
\end{figure}

\end{document}